\documentstyle[titlepage,12pt]{article}

\begin{document}
\title{Chaotic string-capture by black hole}
\vspace{5 in}
\author{A.L.Larsen\thanks{E-mail: allarsen@nbivax.nbi.dk}\\NORDITA,
Blegdamsvej 17, DK-2100 Copenhagen \O, Denmark}
\maketitle
\begin{abstract}
We consider a macroscopic charge-current carrying (cosmic) string in the
background of a Schwarzschild black hole. The string is taken to be circular
and is allowed to oscillate and to propagate in the direction perpendicular
to its plane (that is parallel to the equatorial plane of the black hole).
Nurmerical investigations indicate that the system is
non-integrable, but the interaction with the gravitational field of the
black hole anyway gives rise to various qualitatively simple processes
like "adiabatic capture" and "string transmutation".
\end{abstract}
\newpage
\section{Introduction}
Since Wittens discovery of charge-current carrying topological defects
(cosmic strings) in a $U(1)\times U(\tilde{1})$ gauge theory [1], a large
amount of work has been devoted to the study of their astrophysical and
mathematical aspects. Concerning the mathematical aspects of the
charge-current carrying strings it has been of special interest to consider
the questions of integrability and separability of the equations of motion in
certain electromagnetic and gravitational backgrounds. This seems to be a
natural extension of the work devoted to the study of charged point particles
in such backgrounds.

It is of course well-known that the equations of motion for the Nambu-Goto
string are generally extremely complicated if the string world-sheet is
embedded in a curved 4-dimensional spacetime, and in fact the complete
solution is only known in a very few special cases like conical spacetime [2],
gravitational shock-wave background [3] and a few others. When charges and
currents are introduced on the string the situation is further complicated and
the analysis of the equations of motion for an arbitrary string configuration
in an arbitrary electromagnetic and gravitational background becomes extremely
problematic. A simple way to proceed is to consider only certain families of
string configurations in certain families of backgrounds. Following this
approach Carter, Frolov and their collaborators considered infinitely long
stationary open strings in black hole backgrounds [4,5]. The main results can
be found in Ref. 5:
\vskip 6pt
Starting from a charge-current carrying string described by the action [6]:
\begin{equation}
{\cal S}=\int L(\omega)\sqrt{-\det G_{\alpha\beta}}d\tau d\sigma,
\end{equation}
where:
\begin{equation}
G_{\alpha\beta}=g_{\mu\nu}x^\mu_{,\alpha}x^\nu_{,\beta}
\end{equation}
is the induced
metric on the world-sheet, and $L$ is the Lagrangian taken to be a function
of the world-sheet projection of the gauge covariant derivative of a
world-sheet scalar field $\Phi$ [6]:
\begin{equation}
\omega=G^{\alpha\beta}(\Phi_{,\alpha}+A_\mu x^\mu_{,\alpha})
(\Phi_{,\beta}+A_\mu x^\mu_{,\beta}),
\end{equation}
they made a suitable {\it Ansatz} for the infinitely long stationary open
charge-current carrying string, and were then able to write the equations of
motion in a simple "point particle" Hamiltonian form. This was a considerable
simplification of the problem and they were finally able to show that the
system could be separated for so-called non-dispersive strings in a
Kerr-de Sitter background.

A similar analysis was carried out by the present author [7] for a family of
circular strings. Although the formal calculations were quite similar to
those of Ref. 5, it turned out that the questions of separability and
integrability were less transperent. In this paper we continue the analysis
of the charge-current carrying circular string in a black hole background.
To be more precise we consider the model [8] originally introduced by
Witten [1], which in the notation of Refs. 5 and 6 is obtained by the
choice:
\begin{equation}
L(\omega)=1+\omega/2
\end{equation}
and for simplicity we take a background consisting of the simplest kind
of black hole, namely the Schwarzschild one:
\begin{equation}
ds^2=-(1-2m/r)dt^2+\frac{1}{1-2m/r}dr^2+r^2(d\theta^2+\sin^2\theta d\phi^2).
\end{equation}
We have considered other string models also, for instance the Kaluza-Klein
model of Nielsen [9]. The results in these cases turn out to be quantitatively
similar to the results we will describe below for the model (1.4), also for
more general black hole backgrounds like the Reissner-Nordstr\"{o}m or the
Kerr-Newman black holes [10].

The idea is then to make an {\it Ansatz} that describes charge-current
carrying circular strings and to look at the equations of motion obtained
from the action (1.1) with $L(\omega)$ given by (1.4) and with the
background potentials given by (from (1.5)):
\begin{equation}
A_\mu=0,\hspace*{5mm}g_{\mu\nu}=diag\left( -(1-2m/r),(1-2m/r)^{-1},
r^2,r^2\sin^2\theta\right).
\end{equation}
The circular string in this background will essentially be described by
the same parameters
as a charged point particle and will essentially have the same physical
degrees of freedom. Our analysis will therefore be a natural generalization
of the analysis of the geodesics of point particles in black hole
backgrounds (see for instance Ref. 11). In that spirit
the background (1.6) is considered to
be fixed, i.e. we do not include possible backreactions from the string in our
analysis. Furthermore we do not discuss the question of a possible critical
(maximal) current [8] on the string. We take the more mathematical point
of view and consider both charges and currents as completely arbitrary
quantities.

The paper is organized as follows: In section 2 we consider the equations of
motion for the circular string. We review [7] how a "point particle"
Hamiltonian for the physical degrees of freedom can be obtained, which
simplifies the further analysis considerably. In section 3 we analyse the
effective potential of the Hamiltonian of section 2 with respect to local
and global extrema. In section 4 we continue with some nurmerical
investigations. We consider Poincar\'{e}-plots (surfaces of section) of the
string-dynamics and we
look at different string-trajectories representing different types of
capture and scattering. Finally section 5 contains our conclusions.

We use sign conventions of Misner,Thorne,Wheeler [11] and units in which
beside $G=1, c=1$ the string tension $(2\pi\alpha')^{-1}=1$.
\newpage
\section{The circular string}
\setcounter{equation}{0}
We now consider the circular string in the model described by Eqs. (1.1)-(1.6).
The {\it Ansatz} is:
\begin{equation}
t=t(\tau),\hspace*{5mm}r=r(\tau),\hspace*{5mm}\theta=\theta(\tau),\hspace*{5mm}
\phi=\sigma,
\end{equation}
\begin{equation}
\Phi=f(\tau)+n\sigma,
\end{equation}
where $f$ is an arbitrary function of $\tau$ and $n$ is a constant that is
related to the electromagnetic current on the string [7]. The {\it Ansatz}
(2.1) for the spacetime coordinates describes a plane circular string which
is allowed to oscillate and to propagate in the direction perpendicular to
its plane. For $\theta=\pi/2$ the string is winding around the black hole in
the equatorial plane and for general $\theta$ it is winding around the
$"Z$-axis" keeping its plane parallel to the equatorial plane. The
{\it Ansatz} (2.2) for the scalar field $\Phi$ gives rise to a
uniformly charged and current carrying string [7]. In Ref. 7 it was shown that
after a suitable redefinition of the string time $\tau$ the equations of motion
for the 4 spacetime coordinates can be obtained as Hamilton equations for the
Hamiltonian:
\begin{equation}
H=\frac{1}{2}g^{\mu\nu}P_\mu P_\nu+\frac{1}{2}(r^2\sin^2\theta+n^2+\Omega^2)+
\frac{(\Omega^2-n^2)^2}{8r^2\sin^2\theta},
\end{equation}
with the constraints:
\begin{equation}
H=0,\hspace*{5mm}P_\phi=-n\Omega,
\end{equation}
where $\Omega$ is the constant of motion corresponding to the cyclic coordinate
$\Phi$, and can be interpreted as the constant charge density of the string
[7].

It was already noted in Ref. 7 that because of the $\theta$-dependence of the
potential in Eq. (2.3) the Hamiltonian is not in the Hamilton-Jacobi
separable form [12] (the kinetic energy term is of course separable; it is just
the usual kinetic energy for a massless point particle in the Schwarzschild
background). In principle this could be just a defect of the coordinate system.
Actually the Hamiltonian (2.3) is not even in the separable form in flat
spacetime. In that case the system is however certainly integrable and the
Hamiltonian is in the separable form after transformation to cylinder
coordinates. The same is unfortunately not true in the black hole background
considered here. The point is that when the string is outside the equatorial
plane ($\theta=\pi/2$) it experiences central gravitational forces directed
towards the center of the black hole ($r=0$) as well as non-central forces
from the string tension and the electromagnetic self-interaction directed
towards the center of the circular string ($Z=r\cos\theta\neq 0$). Thus neither
spherical nor cylindrical coordinates for the 3 dimensional space are "good"
coordinates for the dynamical system considered here. In section 4 we will give
nurmerical evidence for the non-integrability of the equations of motion
indicating that the Hamiltonian (2.3) can not be separated in any other
coordinate system neither.

Returning to the form of the Hamiltonian (2.3) we can eliminate the cyclic
coordinates $t$ and $\phi$ by:
\begin{equation}
P_\phi=-n\Omega,\hspace*{5mm}P_t=-{\cal E}_o,
\end{equation}
where ${\cal E}_o$ is the energy of the string. Then we get the 2 dimensional
Hamiltonian:
\begin{equation}
H=\frac{1}{2}(1-2m/r)P^2_r+\frac{1}{2r^2}P^2_\theta+\frac{1}{2}(r\sin\theta+
\frac{N^2}{2r\sin\theta})^2-\frac{{\cal E}^2_o}{2(1-2m/r)},
\end{equation}
where we also introduced the notation:
\begin{equation}
N^2=n^2+\Omega^2.
\end{equation}
Following the point particle case [11] we can write the equation $H=0$ as:
\begin{equation}
\alpha_o{\cal E}^2_o+\gamma_o-(\Delta^2 P^2_r+\Delta P^2_\theta)=0,
\end{equation}
where:
\begin{equation}
\alpha_o=r^4,
\end{equation}
\begin{equation}
\gamma_o=-\Delta(r^2\sin\theta+\frac{N^2}
{2\sin\theta})^2,
\end{equation}
and:
\begin{equation}
\Delta=r^2-2mr.
\end{equation}
We can then define an effective potential $U(r,\theta)$ [11]:
\begin{equation}
\alpha_o U^2(r,\theta)+\gamma_o=0.
\end{equation}
If $U(r,\theta)={\cal E}_o$ we have a turning point ($P_r=P_\theta=0$). The
string, which is now described by the point particle coordinates $r$ and
$\theta$, is then restricted to the area where $U(r,\theta)\leq{\cal E}_o$.
The explicit expression for $U(r,\theta)$ is:
\begin{equation}
U(r,\theta)=\sqrt{\Delta}(\sin\theta+\frac{N^2}{2r^2\sin\theta}).
\end{equation}
In the next section we will analyse this potential in some detail. It is
then convenient also to express it in terms of cylinder coordinates
($R=r\sin\theta,\hspace*{2mm}Z=r\cos\theta$):
\begin{equation}
\tilde{U}(R,Z)=(R+\frac{N^2}{2R})\sqrt{1-\frac{2m}{\sqrt{R^2+Z^2}}}.
\end{equation}
In these coordinates $R$ is the radius of the string loop and $\mid Z\mid$
measures the distance from the equatorial plane of the black hole to the
plane of the string. Note that $\tilde{U}(2m,0)=0$, i.e. the global minimum
of the potential outside the horizon of the black hole is at the horizon in
the equatorial plane.
\newpage
\section{The effective potential}
\setcounter{equation}{0}
In this section we analyse the potential (2.14) in more detail. A typical
picture of the potential is shown in Fig.1. We see that the potential is
essentially an elongated valley between 2 mountain chains. The mountain
chain to the right (larger $R$) represents the string tension trying to
collapse the circular string, while the mountain chain to the left (smaller
$R$) represents the electromagnetic self-interactions trying to expand the
circular string. The most interesting observation is however that near the
equatorial plane the gravitational attraction between the black hole and the
string may overcome the other forces involved. This gives rise to a mountain
pass from the valley through the horizon into the black hole. The only chance
for the charged and/or current carrying circular string to collapse is
therefore to go through this mountain pass and to fall into the black hole.

Let us now consider the critical points of the potential:
\begin{equation}
\frac{\partial\tilde{U}}{\partial R}=(1-\frac{N^2}{2R^2})(1-\frac{2m}{\sqrt{
R^2+Z^2}})^{1/2}+\frac{mR}{(R^2+Z^2)^{3/2}}(R+\frac{N^2}{2R})(1-\frac{2m}{
\sqrt{R^2+Z^2}})^{-1/2}=0,
\end{equation}
\begin{equation}
\frac{\partial\tilde{U}}{\partial Z}=\frac{mZ}{(R^2+Z^2)^{3/2}}(R+\frac{N^2}
{2R})(1-\frac{2m}{\sqrt{R^2+Z^2}})^{-1/2}=0.
\end{equation}
Since we are looking for solutions outside the horizon ($R^2+Z^2=r^2\geq 4m^2$)
we find that necessarily $Z=0$. It follows that $R=r$ is determined by:
\begin{equation}
r^3-mr^2-\frac{N^2}{2}r+\frac{3}{2}mN^2=0.
\end{equation}
The second derivatives are given by:
\begin{equation}
\frac{\partial^2\tilde{U}}{\partial R\partial Z}\mid_{crit.}=0,
\end{equation}
\begin{equation}
\frac{\partial^2\tilde{U}}{\partial Z^2}\mid_{crit.}=
\frac{m(r+\frac{N^2}{2r})}{r^2\sqrt{r^2-2mr}}\mid_{crit.}>0,
\end{equation}
\begin{equation}
\frac{\partial^2\tilde{U}}{\partial R^2}\mid_{crit.}=
\frac{1}{r^2 (r^2-2mr)^{3/2}}\left( (N^2-m^2)r^2-6mN^2r+\frac{15}{2}m^2N^2
\right)\mid_{crit.}
\end{equation}
So the equatorial plane is a stable plane and the critical points obtained
as solutions to Eq. (3.3) are stable (unstable) if (3.6) is positive
(negative). In the case illustrated in Fig.1. there is obviously both a
stable and an unstable critical point outside the horizon. This is actually
the most interesting case since we can then have a stationary circular string
winding around the black hole in the equatorial plane [10]. By carefully
analysing the solutions of the cubic equation (3.3) and the corresponding
signs of expression (3.6) we find that for:
\begin{equation}
N^2>(13\sqrt{13}+47)m^2\equiv N^2_{crit.}
\end{equation}
there is a stable and an unstable critical point outside the horizon (Fig.1.).
If $N^2<N^2_{crit.}$ there are no critical points outside the horizon and
therefore a string in the equatorial plane will always collapse and fall into
the black hole. Finally for $N^2=N^2_{crit.}$ we get the (unstable) critical
point closests to the horizon:
\begin{equation}
\frac{\partial^2\tilde{U}}{\partial R^2}\mid_{crit.}=\frac{\partial\tilde{U}}{
\partial R}\mid_{crit.}=0,
\end{equation}
with solution:
\begin{equation}
N^2=N^2_{crit.},\hspace*{5mm}r_{crit.}=\frac{1}{2}(5+\sqrt{13})m,
\end{equation}
i.e. whatever the charges and currents are we can never have stationary
strings with $r\leq r_{crit.}$ winding around the black hole.

In the rest of this paper we will only consider the case (3.7). So we have
a global minimum outside the horizon at $(R,Z)\equiv (r_+,0)$ and a
mountain pass (a saddle point) from the valley towards the black hole at
$(R,Z)\equiv (r_-,0)$:
\begin{eqnarray}
&\frac{\partial\tilde{U}}{\partial R}\mid_{R=r_+, Z=0}=
\frac{\partial\tilde{U}}{\partial R}\mid_{R=r_-, Z=0}=0,&\nonumber\\
&\frac{\partial^2\tilde{U}}{\partial R^2}\mid_{R=r_+, Z=0}>0,\hspace*{5mm}
\frac{\partial^2\tilde{U}}{\partial R^2}\mid_{R=r_-, Z=0}<0,&\nonumber\\
&\frac{\partial^2\tilde{U}}{\partial Z^2}\mid_{R=r_\pm, Z=0}>0,\hspace*{5mm}
\frac{\partial^2\tilde{U}}{\partial R\partial Z}\mid_{R=r_\pm, Z=0}=0.&
\nonumber
\end{eqnarray}
{}From equation (3.3) we find:
\begin{equation}
r_+=\frac{m}{3}\left(1+\sqrt{4+6\frac{N^2}{m^2}}\cos(\frac{\psi}{3}+
\frac{4\pi}{3})\right),
\end{equation}
\begin{equation}
r_-=\frac{m}{3}\left(1+\sqrt{4+6\frac{N^2}{m^2}}\cos\frac{\psi}{3}\right),
\end{equation}
where:
\begin{equation}
\cos\psi=\frac{\sqrt{8}m(m^2-18N^2)}{(2m^2+3N^2)^{3/2}}\in[-1,0[.
\end{equation}

We close this section by considering strings infinitely far away from the black
hole. The mountain chain due to the tension of the string becomes infinitely
high:
\begin{equation}
\tilde{U}(R,Z)\rightarrow\infty,\hspace*{5mm}for\hspace*{5mm}
R\rightarrow\infty,
\end{equation}
so that the only way for a string with finite energy to escape from the
black hole is to go in the $Z$-direction:
\begin{equation}
\tilde{U}(R,\pm\infty)=R+\frac{N^2}{2R}.
\end{equation}
A stationary circular configuration at $Z=\pm\infty$ will be in the minimum
of this potential, which is situated at $R=N\sqrt{2}$. The corresponding
potential energy is:
\begin{equation}
\tilde{U}(\frac{N}{\sqrt{2}})=\sqrt{2}N.
\end{equation}
It is interesting to compare the potential energy (3.15) with the
corresponding energies at the points $(r_\pm,0)$ given in Eqs. (3.10)-(3.11).
Obviously:
\begin{equation}
\tilde{U}(\frac{N}{\sqrt{2}},\pm\infty)>\tilde{U}(r_+,0),
\end{equation}
\begin{equation}
\tilde{U}(r_-,0)>\tilde{U}(r_+,0),
\end{equation}
and nurmerically we find:
\begin{equation}
\tilde{U}(\frac{N}{\sqrt{2}},\pm\infty)\geq\tilde{U}(r_-,0)\hspace*{5mm}
\Longleftrightarrow\hspace*{5mm}\mid N\mid\leq (13.22...)m.
\end{equation}
We will return to these conditions in the following section.
\section{Nurmerical solutions}
\setcounter{equation}{0}
We now return to the equations of motion corresponding to the 2 dimensional
Hamiltonian (2.6). In explicit form they read:
\begin{equation}
\ddot{\theta}=-\frac{2\dot{r}\dot{\theta}}{r}-\sin\theta\cos\theta(1-
\frac{N^4}{4r^4\sin^4\theta}),
\end{equation}
\begin{equation}
\ddot{r}=(r-3m)\dot{\theta}^2-(r-m)\sin^2\theta-\frac{mN^2}{r^2}+
\frac{(r-3m)N^4}{4r^4\sin^2\theta}.
\end{equation}
{}From the constraint $H=0$ we have the first integral:
\begin{equation}
{\cal E}^2_o=\dot{r}^2+(r^2-2mr)\dot{\theta}^2+(r^2-2mr)(\sin\theta+
\frac{N^2}{2r^2\sin\theta})^2.
\end{equation}
Unfortunately we have not been able to find an other first integral, necessary
to separate Eqs. (4.1)-(4.2). We will therefore turn to nurmerical methods
and in that proces we will actually provide some evidence
that no second first integral exists. To integrate Eqs. (4.1)-(4.2)
nurmerically it is convenient to introduce the dimensionless quantities
$(x,y,\bar{N})$:
\begin{equation}
r\equiv xm,\hspace*{5mm}\theta\equiv y,\hspace*{5mm}N\equiv\bar{N}m,
\end{equation}
and to write the equations of motion in first order form:
\begin{equation}
\dot{x}=p,
\end{equation}
\begin{equation}
\dot{y}=q,
\end{equation}
\begin{equation}
\dot{p}=(x-3)q^2-(x-1)\sin^2y-\frac{\bar{N}^2}{x^2}+\frac{(x-3)\bar{N}^4}{
4x^4\sin^2 y},
\end{equation}
\begin{equation}
\dot{q}=-\frac{2pq}{x}-\sin y\cos y (1-\frac{\bar{N}^4}{4x^4\sin^4 y}).
\end{equation}
We have integrated this system of equations using the fourth order
Runge-Kutta method for a variety of initial data, of which we now present
a few.

First let us consider the situation where the string is initially at rest
(non-propagating and non-oscillating) at $Z=-\infty$, i.e. it is described
by Eq. (3.15). Because of the gravitational interaction the string will start
moving towards the equatorial plane of the black hole. Since the string is
coming from $Z=-\infty$ it apriory has enough energy to escape to $Z=+\infty$
after having passed the black hole. However, this will generally not happen.
The point is that part of the translational energy of the c.o.m. of the string
is transformed into oscillational energy, and the string will therefore be
"trapped" in the vicinity of the equatorial plane. This is a typical
"adiabatic invariance" phenomenon [12] for motion in a 2 dimensional elongated
potential. This is shown in Fig.2a. There are now 2 possibilities depending
on the charges and currents on the string. If $\mid N\mid\leq(13.22...)m$ the
string has enough energy to pass the mountain pass towards the center of the
black hole (see Eq. (3.18)). After "jumping" around the equatorial plane
for some time the
string will eventually hit the mountain pass and collapse into the black hole.
If on the other hand $\mid N\mid>(13.22...)m$ the string will be "adiabatically
trapped" around the equatorial plane forever (Fig.2a),
but will never actually fall into
the center of the black hole. In the latter case it is convenient
to consider the surface of section defined by:
\begin{equation}
\theta=\pi/2,\hspace*{5mm}\dot{\theta}\geq 0,
\end{equation}
and to plot $(\dot{R},R)$ (so-called Poincar\'{e} plots) for various values
of $N$. A typical plot is shown in Fig.2b. for $\bar{N}=14$. The result is a
completely irregular collection of points, that strongly indicates that no
second integral of motion exists for the system (4.5)-(4.8), i.e. it is
non-integrable (compare with similar plots for say the H\'{e}non-Heiles
system [13]).

The "adiabatic capture" processes described above are obviously most
relevant for strings initially at rest. If the string is instead in an
initial state of oscillation and/or propagation the typical picture is
that the interaction with the black hole gives rise to a change in
the distribution of translational and oscillational energy, i.e. if the string
is in one oscillating state at $Z=-\infty$ it will move towards the
equatorial plane, pass the black hole and continue towards $Z=+\infty$ in
another oscillating state. Such processes we denote as "string transmutation"
and 2 examples are shown in Fig.3. Finally it
is interesting to remark
that similar processes were found in a completely different context by de Vega
and S\'{a}nchez [14]. They considered the quantum scattering of microscopic
fundamental
strings on a black hole whereas we consider classical scattering of
macroscopic charge-current carrying circular cosmic strings on a black hole.
Note that in describing these processes
we have neglected such phenomena as electromagnetic and gravitational radiation
from the oscillating string. As explained in the Introduction treating such
effects is out of the scope of this paper.

\section{Conclusion}
\setcounter{equation}{0}
In this paper we continued our investigations of charge-current
carrying circular strings in
black hole backgrounds [7,10]. Following the approach of Ref. 7 we
obtained a 2 dimensional "point particle" Hamiltonian (2.6) determining the
dynamics of the string. From this Hamiltonian we defined an effective
potential which was analysed with respect to local and global extrema in
section 3. An important result of this analysis was the discovery of
a critical charge-current (3.7) implying the existence of stable
stationary strings winding around the black hole in the equatorial plane.
Nurmerical investigations in section 4 indicated that, in contrary to the
somewhat similar point particle case, the equations of motion (4.5)-(4.8) are
non-integrable and we gave an example where the string is in fact jumping
chaotically around the equatorial plane of the black hole (Fig.2.). Finally
we presented different kinds of string trajectories representing "adiabatic
capture" and "string transmutation" processes.
\vskip 12pt
\hspace*{-6mm}{\bf Acknowledgements}:
I would like to thank J. J\o rgensen for helping me with the nurmerical
calculations described in section 4, and S.E. Rugh for discussions on
various aspects of chaos, integrability and solvability.
\newpage
\begin{centerline}
{\bf References}
\end{centerline}
\begin{enumerate}
\item Witten E 1985 {\it Nucl. Phys.} {\bf B249} 557
\item de Vega H J, Ramon-Medrano M and S\'{a}nchez N 1992 {\it Nucl. Phys.}
      {\bf B374} 405
\item de Vega H J and S\'{a}nchez N 1989 {\it Nucl. Phys.} {\bf B317} 706, 731
\item Frolov V P, Skarzhinsky V D, Zelnikov A I and Heinrich O 1989
      {\it Phys. Lett.} {\bf B224} 255, Carter B and Frolov V P 1989
      {\it Class. Quantum Grav.} {\bf 6} 569, Carter B 1990 {\it Class.
      Quantum Grav} {\bf 7} L69
\item Carter B, Frolov V P and Heinrich O 1991 {\it Class. Quantum Grav.}
      {\bf 8} 135
\item Carter B 1989 {\it Phys. Lett.} {\bf B224} 61
\item Larsen A L 1993 {\it Class. Quantum Grav.} To appear
\item Copeland E, Hindmarsh M and Turok N 1987 {\it Phys. Rev. Lett.}
      {\bf 58} 1910, Vilenkin A and Vachaspati T 1987 {\it Phys. Rev. Lett}
      {\bf 58} 1041
\item Nielsen N K 1980 {\it Nucl. Phys.} {\bf B167} 249
\item Larsen A L 1991 {\it Phys. Lett.} {\bf B273} 375
\item Misner C W, Thorne K S and Wheeler J A 1973 Gravitation ({\it Freeman,
      San Francisco, CA})
\item Goldstein H 1980 Classical Mechanics 2nd. edition ({\it New York,
      Addison-Wesley}) 531
\item H\'{e}non M and Heiles C 1964 {\it Astronom. J.} {\bf 69} 73
\item de Vega H J and S\'{a}nchez N 1988 {\it Nucl. Phys.} {\bf B309} 552, 577
\end{enumerate}
\newpage
\begin{centerline}
{\bf Figure captions}
\end{centerline}
\hspace*{-6mm}Fig.1. The effective potential $\tilde{U}(R,Z)$. In the
equatorial plane ($Z=0$) we find the global minimum and the mountain pass
connecting the elongated valley and the center of the black hole.
\vskip 12 pt
\hspace*{-6mm}Fig.2. "Adiabatic capture". ($R,Z$)-trajectory (a)
for a string with $\bar{N}=14$
that is initially in a stationary state at $Z=-\infty$, and the
corresponding Poincar\'{e} plot (b) indicating non-integrability of the
system (4.5)-(4.8).
\vskip 12 pt
\hspace*{-6mm}Fig.3. "String transmutation". ($R,Z$)-trajectories
for strings that are
initially in oscillating but non-propagating states at ($Z=-\infty$). The
gravitational interaction
with the black hole changes the state of oscillation.
\end{document}